\documentclass[%
aip,
rsi,%
amsmath,amssymb,
reprint,%
]{revtex4-1}

\usepackage{graphicx}
\usepackage{dcolumn}
\usepackage{bm}

\begin{document}
	
	\title{Additive 3D photonic integration that is CMOS compatible}
	
	\author{Adria Grabulosa}
	\affiliation{FEMTO-ST Institute/Optics Department, CNRS \& University Franche-Comt\'e, \\15B avenue des Montboucons,
	Besan\c con Cedex, 25030, France}

	\author{Johnny Moughames}
\affiliation{FEMTO-ST Institute/Optics Department, CNRS \& University Franche-Comt\'e, \\15B avenue des Montboucons,
	Besan\c con Cedex, 25030, France}

	\author{Xavier Porte}
\affiliation{FEMTO-ST Institute/Optics Department, CNRS \& University Franche-Comt\'e, \\15B avenue des Montboucons,
	Besan\c con Cedex, 25030, France}
\affiliation{Now with: Institute of Photonics, Department of Physics, University of Strathclyde, Glasgow G1 1RD, UK}

	\author{Muamer Kadic}
\affiliation{FEMTO-ST Institute/Optics Department, CNRS \& University Franche-Comt\'e, \\15B avenue des Montboucons,
	Besan\c con Cedex, 25030, France}

	\author{Daniel Brunner}
	\email{daniel.brunnerfemto-st.fr}
	\affiliation{FEMTO-ST Institute/Optics Department, CNRS \& University Franche-Comt\'e, \\15B avenue des Montboucons,
		Besan\c con Cedex, 25030, France}

	\date{\today}
	
	\begin{abstract}
		
Today, continued miniaturization in electronic integrated circuits (ICs) appears to have reached its fundamental limit at $\sim$2~nm feature-sizes, from originally $\sim$1~cm.
At the same time, energy consumption due by communication becomes the dominant limitation in high performance electronic ICs for computing, and modern computing concepts such a neural networks further amplify the challenge.
Communication based on co-integrated photonic circuits is a promising strategy to address the second.
As feature size has leveled out, adding a third dimension to the predominantly two dimensional integrated circuits appears the most promising future strategy for further IC architecture improvement.
Crucial for efficient electronic-photonic co-integration is CMOS compatibility of the associated photonic integration fabrication process.
Here, we review our latest results obtained in the FEMTO-ST RENATECH facilities on using additive photo-induced polymerization of a standard photo-resin for truly 3D photonic integration according to these principles.
Based on one- and two-photon polymerization and combined with direct-laser writing, we 3D-printed air- and polymer-cladded photonic waveguides.
An important application of such circuits are the interconnects of optical neural networks, where 3D integration enables scalability in terms of network size versus its geometric dimensions.
In particular via \emph{flash}-TPP, a fabrication process combining blanket one- and high-resolution two-photon polymerization, we demonstrated polymer-cladded step-index waveguides with up to 6~mm length, low insertion ($\sim$0.26~dB) and propagation ($\sim$1.3~dB/mm) losses, realized broadband and low loss ($\sim$0.06~dB splitting losses) adiabatic 1 to M couplers as well as tightly confining air-cladded waveguides for denser integration.
By stably printing such integrated photonic circuits on standard semiconductor samples, we show the concept's CMOS compatibility.
With this, we lay out a promising, future avenue for scalable integration of hybrid photonic and electronic components.

	\end{abstract}
	
	\maketitle

\section{Introduction}

The backbone behind most of today's cutting-edge technology is dense integration of two dimensional (2D) electronic circuits.
However, by now these do experience several challenges.
Further pushing the performance of 2D computing chips becomes increasingly difficult, while new applications, in particular neural networks (NNs), challenge the hegemony of such 2D circuits - and this on a fundamental level \cite{Dinc2020, Boahen2022}.
New integration concepts and fabrication technologies are needed if we are to continue the astonishing technological progress of the past decades.
Crucially, these integration concepts need to take the essential features behind the success of 2D electronic integrated circuits (ICs) into account.

Elevating a new integration technology even close to the level of 2D electronic ICs is a daunting and certainly a long-term challenge.
Since the first demonstration of a planar, i.e. 2D, monolithic IC at Fairchild, this classical integration has continuously been advanced for 60 years plus in an almost world-wide effort.
The concept's success is a testimony to what can be achieved when previously individual components are integrated inside a single, monolithic circuit.
It typically led to substantial miniaturization and increased reliability as well as robustness, all while fabrication costs plummeted.
Combined, these factors enabled decades of exponential scaling for electronic ICs: around every two years the numbers of transistors per chips doubled (Moore's law) while the power consumption per component halves (Dennard scaling).
Monolithic ICs comprising different components and functionalities are therefore also indispensable for 3D photonic integration.

While still far from the levels of today's electronic IC, photonic integration also has considerably advanced.
In order to maximize compatibility and synergy with electronics, photonic integration based on silicon substrates emerged in the 1980s with the demonstration of the silicon waveguide \cite{Boyd1985,Soref1986}, the photonic equivalent to a metallic or polysilicon  wire in integrated electronics ICs.
Electronic ICs are almost exclusively based on complementary metal–oxide–semiconductor (CMOS) technology that uses mostly silicon as semiconductor host leveraging boron, gallium, indium, phosphorus, arsenic and bismuth as dopants, and CMOS compatibility is considered fundamentally important for photonic ICs.

By a vast majority, both, electronic and photonic integration leverages fabrication concepts developed for planar, 2D substrates.
The layout of a circuit's single layer is etched into a thin surface of either mostly metal or semiconductor materials, which is the process of 2D lithography.
Typically, coating said surface with a photo-resist protects certain surface-areas from etching, which is determined by photo-resist illumination that is structured by a photo-mask. 
The appeal of such 2D lithography is that each of the involved process steps, photo-resist application, exposure by photo-mask, etching and several washing sequences, can be carried out in a single procedure for a large area or even an entire wafer, which strongly reduces fabrication costs.

\begin{figure}[t]
	\centering
	\includegraphics[width=0.8\linewidth]{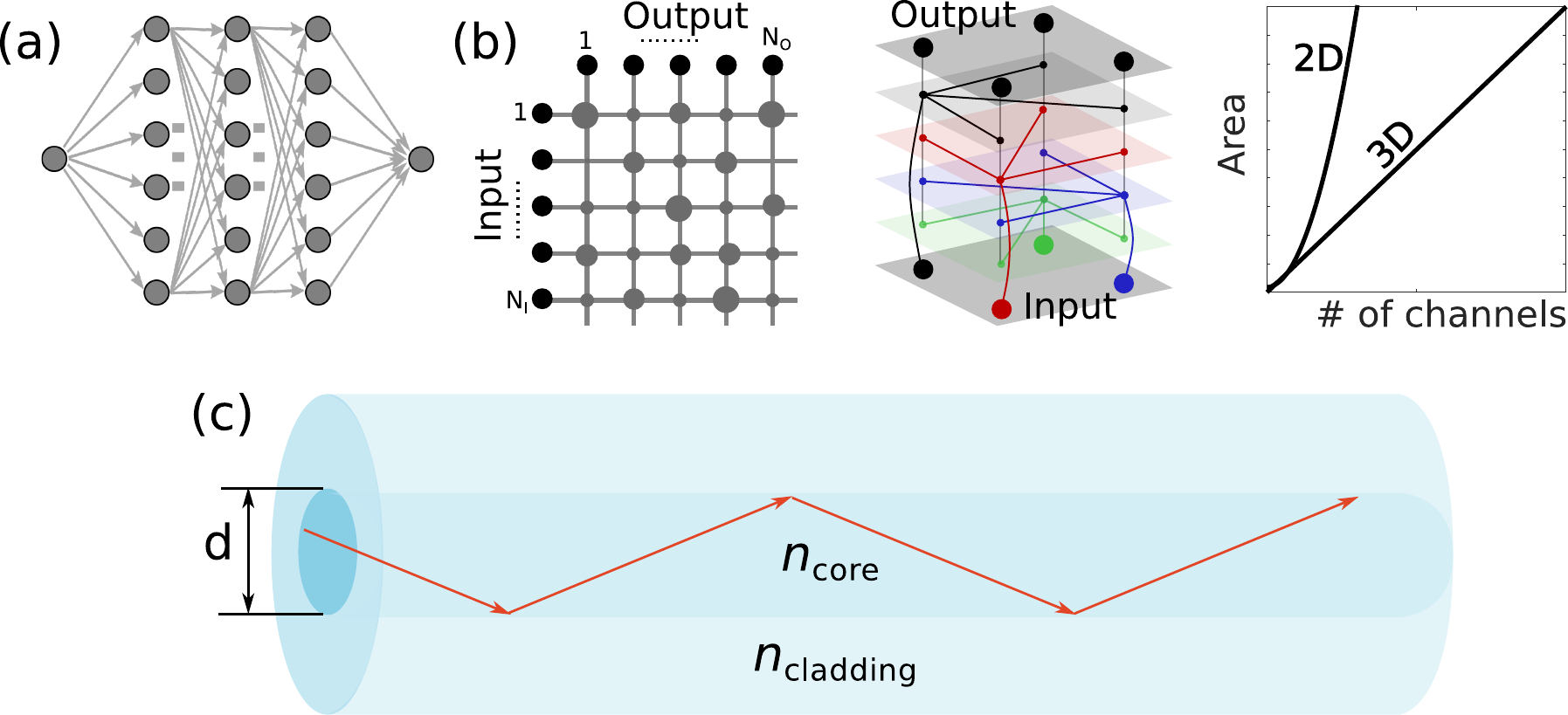}
	\caption{3D photonic integration and optical waveguide basics.
		(a) Schematics of a typical neural network where a large number of neurons are highly interconnected through a network. 
		(b) Integrating a large number connections in 2D leads to an exponential growth of the number of channels over the chip's area; whereas leveraging integration in 3D results in a efficient and linear scalability of optical interconnects. 
		(c) In photonic waveguides, the light is confined within the core of diameter $d$ due to total internal reflection.
		For this, the refractive index of the core $n_{\textrm{core}}$ must be larger than the cladding's $n_{\textrm{cladding}}$, and hence $\Delta{n}=n_{\textrm{core}}- n_{\textrm{cladding}}>0$. 
		All the waveguide's optical properties relies on the parameters $\Delta{n}$ and $d$.
		\label{fig1}}
\end{figure}

A new challenge to classical electronics computers based on 2D substrates arose with the breakthrough of NN computing around a decade ago.
Conceptually, NNs link a large number of neurons through the network's connections, c.f. Fig.~\ref{fig1} (a).
In an physical hardware implementation that mirrors this topology, these connections correspond to electronic or photonic signaling 'wires'.
Currently, these connections are emulated, which creates substantial energy and speed overheads.
Future NN circuits that abolish this overhead require ICs with a far higher degree of connectivity, i.e. much more wires to communicate signals across the chip.
This causes several problems.
Energetically speaking, electronic communication is the factor limiting performance even for classical computing concepts; communicating a floating point number costs around 80-times more energy than creating a new floating point number \cite{Dally2018}.
NN computation dramatically escalates this problem, as the number of a NN's connections by far out-scale the number of neurons.
Photonic and 3D integration provide promising solutions, see Fig.~\ref{fig1} (b).
Optical communication is (i) energetically superior for ever shorter distances and (ii) mitigates heat dissipation challenges that arise for volumetric circuits, while (iii) 3D integration shortens the length of communication links.
Most importantly, in many NN topologies the number of connections, i.e. wires, increases quadratic or faster with the number of neurons.
Consequently, integrating a NN's interconnect in 2D results in a quadratic scaling (or worse) of chip-area with the size of a neural network.
Recently, the number of neurons in a NN has turned into the parameter of fundamental relevance, and alternative strategies for integrating NNs are of fundamental importance for the field.

In this review for the RENATECH special issue, we describe our recent work addressing such photonic ICs based on standard techniques and fabrication infrastructure available in our local RENATECH cleanroom.
In those efforts, we have demonstrated additive, 3D photonic integration, which importantly is using concepts and materials that make the entire fabrication and resulting photonic IC CMOS compatible.
Based on additive two-photon polymerization (TPP) in a direct-laser writing (DLW) system, combined with rapid and large area one-photon polymerization (OPP), we integrated large 3D photonic waveguide circuits.
We demonstrate individual waveguides as well as optical splitters and networks of splitter \cite{Grabulosa2023} based on (i) air-cladded waveguides comprising polymer cores \cite{Moughames2020}, and (ii) step-index waveguides where we induce the refractive index difference between core and cladding required for guiding by dynamically controlling the optical power used for printing our 3D structures \cite{Porte2021}.
Finally, we substantially accelerate the fabrication process by developing the \emph{flash}-TPP concept, which combines TPP-DLW with ultraviolet (UV) blanked illumination to efficiently polymerize an IC's non-light guiding volume in a single step \cite{Grabulosa2022}.
We achieve very symmetric splitting ratios in optical couplers, and (for a first proof of concept) low propagation losses of $\sim1.3~$dB/mm and insertion losses of $\sim0.26~$dB.
Finally, we printed optical waveguides on semiconductor substrates hosting micro-lasers, demonstrating that our concept is CMOS compatible.

\section{\label{sec:DLW} Basics of additive fabrication}

In the past 15 years, DLW and TPP have become a versatile fabrication tool of polymer structures with sub-micron dimensions \cite{deubel2004direct,moughames2016wavelength,anscombe2010direct}.
In contrast to 2D planar methods such as electron‐beam lithography or mask based lithography, DLW allows for fabricating three‐dimensional structures \cite{wang20223d}.
DLW has played a crucial role for many proof-of-concept designs in optics \cite{Moughames2020}, acoustics \cite{iglesias2021three,frenzel2019ultrasound}, elasticity \cite{chen20223d,chen2022closed,wang20223d,dudek2022micro}, robotics \cite{ji20214d} and even electric transport \cite{kern2017experimental}.
Major challenges such as inclusion of conductive resins \cite{Blasco2016}, quantum-dots doped resins \cite{Mayer2017}, liquid-crystals doped resins \cite{munchinger20223d} are still in the development phase.
Recently, great progress towards parallel direct-laser writing has been made, which enables a substantially accelerated fabrication process \cite{kiefer2022parallelizing}.
Finally, different polymerization concepts are constantly being developed, some of which use novel approaches to high-resolution 3D printing based on polymer resins \cite{hahn2022light}.\\ 

\section{\label{sec:polymerinteg} Photonic integration via photo-induced polymerization}

Standard photonic waveguides covered in this review rely the guiding element called the core having a higher refractive index $n_{\textrm{core}}$ than the refractive index of the confining part called the cladding $n_{\textrm{cladding}}$, i.e. $\Delta{n}=n_{\textrm{core}}- n_{\textrm{cladding}}>0$.
As schematically illustrated in Fig.~\ref{fig1} (c), in such a configuration optical rays impinging on the core-cladding interface with an angle smaller than the critical angle $\theta_{\textrm{c}}=\arcsin(1-(\Delta{n}/n_{\textrm{core}}))$ exhibit total internal reflection.
As a consequence, they are confined to the waveguide's core and propagate along this structure, allowing to direct optical propagation along  pre-designed paths via an integrated and solid core.

Refractive index contrast $\Delta{n}$ combined with the core diameter $d$ are a waveguide's determining characteristics, which determine a waveguide's numerical aperture $\textrm{NA}=\sqrt{n_{\textrm{core}}^2- n_{\textrm{cladding}}^2}$.
The same holds for the number of spatial modes allowed to propagated through the waveguide $M\approx{V}^2 / 2 = (4\pi{d} / \lambda)\textrm{NA}$ for large $M$, where $\lambda$ is the optical wavelength..
Here, $V$ is the normalized frequency a central indirect property of optical waveguides; for $V\leq{2.405}$ a waveguide is single-mode, otherwise it allows for higher modes to propagate.
Finally, $\Delta{n}$ also determines the minimal bending radius for which light can be directed without exceedingly high losses.
This in turn is the limiting factor for integration density inside a photonic IC.

In work covered in this review, we used the commercial 3D direct-laser writing Nanoscribe GmbH (Photonics Professional GT) system, which is equipped with a femtosecond (fs) laser operating at 780~nm, and galvo-mirrors for rapid beam movement in the lateral directions. 
The fs-laser is usually tightly focused into the resin through an objective lens of high numerical aperture.
After finishing the TPP-DLW step, the unpolymerized resin was removed in a two-step development process, immersing the structure first in propylene-glycol-methyl-ether-acetate (PGMEA) acting as a developer for 20 minutes, followed by rinsing in isopropyl alcohol (2-propanol) for 3-5 minutes.
For OPP, we deposited samples in the commercial UV-chamber Rolence Enterprise Inc., LQ-Box model, 405 nm wavelength, 150 mW/cm$^2$ average light intensity.

\subsection{\label{subsec:TPP}Two-photon polymerization}

\begin{figure}[h!]
	\centering
	\includegraphics[width=0.95\linewidth]{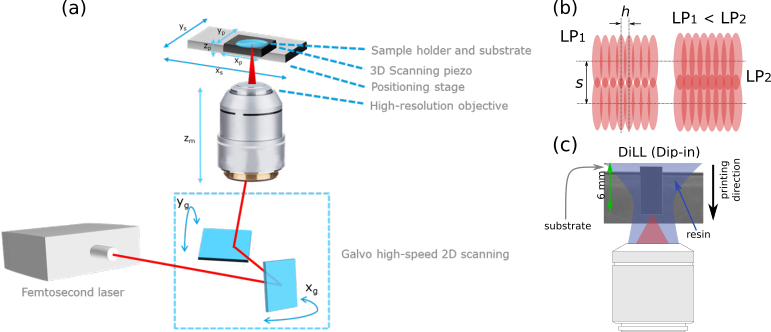}
	\caption{Principle of direct-laser writing (DLW). (a) The fs-writing laser is scanned through the photo-resist through the monomer resin using high-speed galvo-mirrors for the displacement in the $(x,y)$-plane, while a piezo controls the $z$-position.
		(b) The resin is two-photon polymerized only inside a small voxel volume, and voxels are placed on a grid determined by hatching distance $h$ in the $(x,y)$-plane, and slicing distance $s$ in the z-direction.
		The laser power (LP) as well as $s,h$ determine the overlap of neighboring voxels and through this the minimum feature size and the smoothness of printed surfaces.
		(c) In our work we use the 'dip-in' technique, where a drop of resin is located between the microscope objective and the substrate.
		The printing direction is downwards, and the maximum size of 3D-printed structures is around 6~mm in height.\label{fig2}}
\end{figure}

Two-photon polymerization is a maskless direct-laser writing technique~\cite{Hong2004}.
A highly focused pulsed laser beam in the femtosecond regime is used to induce the absorption of two-photons in the exposed volume inside the photo-resist (which is a monomer in its liquid phase), c.f Fig.~\ref{fig2} (a).
This two-photon activated polymerization creates long-chained polymer molecules that in turn form a solid volume due to molecule interlinkage.
Forming almost arbitrary 3D structures can then be achieved by translating the laser through the overall volume of the photo-resist along all three spatial dimensions.
Gravity can impose limitations on attainable shapes, yet this aspect usually does not have a too strong impact: the polymer and the original monomer resin have very similar mass densities, and thus the Archimedes forces keep a polymerized voxel locked in its position due to the resin's viscosity.

Originally, the writing laser spot was translated through the resin using piezo stages.
This approach is highly accurate as the stages readily have nanometric precision.
However, it does not allow for large displacement, is very slow and hence cannot be used for large printing areas/volumes.
A major breakthrough resulted from using galvo-mirrors for moving the writing laser's focal spot through the resin (see Fig.~\ref{fig2} (a)).
As a consequence, printing speed increased by orders of magnitude \cite{buckmann2014three}, and fabricating large-scale 2.5 metasurfaces or 3D volumes became possible. 

Crucial for the quality of 3D structures and for integration in general is the feature size of a single, polymerized voxel relative to the the scanning speed of the printing laser.
The photoinitiation of the chemical reaction which essentially is instantaneous relative to the the writing speed, and hence the writing-volume directly follows laser's scanning.
However, polymerization is a chemical reaction with an associated time scale, like any diffusion phenomenon.
Typically, this timescale is orders of magnitude slower than the galvo-controlled laser scanning \cite{yang2019schwarzschild}.
This aspect is crucial, since as a consequence polymerization is taking place for several neighboring voxels at overlapping times.
It makes the polymerization process more homogeneous, and the obtained structures do not suffer from (unintended) variations of material properties resulting from stitching  countless small voxels together to form a large structure.
As schematically illustrated in Fig.~\ref{fig2} (b), the writing laser power (LP), the hatching $h$ and slicing $s$ distances as well as the scan speed modify the overlap between neighboring polymer voxels.
Through this, the smoothness of surfaces and the homogeinity of the polymer-medium can  be controlled to a good degree.
For much faster polymerization, the periodic voxels would results in a photonic crystal like structure, thus introduce scattering and all related phenomena inside the produced polymer.
Thanks to diffusion, this aspect is almost not observable, yet it potentially is a source of optical losses in long waveguides. 

A powerful technique, called 'dip-in' mode, c.f. Fig.~\ref{fig2} (c), where the liquid resin is held between the substrate and the microscope objective, was introduced in 2013.
This avoids having to print through the substrate (contrary to immersion-oil techniques), which reduces aberrations and removes the thickness of the substrate as a limitation of the maximal height of printed structures.
Importantly for CMOS compatibility, it enables printing on materials that are not transparent at fs-laser's wavelength.
Piezo actuators and/or the writing field (determined by the microscope objective of the printer) are usually quite limited in area, usually below mm-scales.
For printing larger structures stitching various writing fields together is required, and in that it is not dissimilar to the \emph{stepper}-process used in 2D semiconductor lithography.
One can select a lower NA microscope objective to increase the writing field, however, this can only be employed on the cost of a reduced printing low-resolution \cite{Ristok2020}.

Generally, 3D printing via direct-laser writing creates structures of high quality, and their optical and ellastical properties have been characterized with high accuracy using Brillouin light scattering \cite{ugarak2022brillouin}.
In this paper, the authors demonstrate an excellent quality check of the polymer in the GHz regime for elastic waves.
For example, the 3D-printed samples can have an elastic quality factor only ten times smaller than fused silica at hypersonic frequencies.

Importantly for printing photonic waveguides, the degree of polymerization and through the Clausius relationship also the refractive index $n$, is mainly determined by the type of photo-resist and the dose parameters $D$ of the fs-laser, i.e. scanning speed and LP. 
Within the window between the TPP-threshold and the breakdown point above which the polymerized voxel contains defects, the so-called dynamic power range of the photo-resist~\cite{Hong2004}, the size of the TPP-voxel can be further modified by adapting $D$ and fabrication parameters distances $h$ and $s$.

\begin{figure}[h!]
	\centering
	\includegraphics[width=0.95\linewidth]{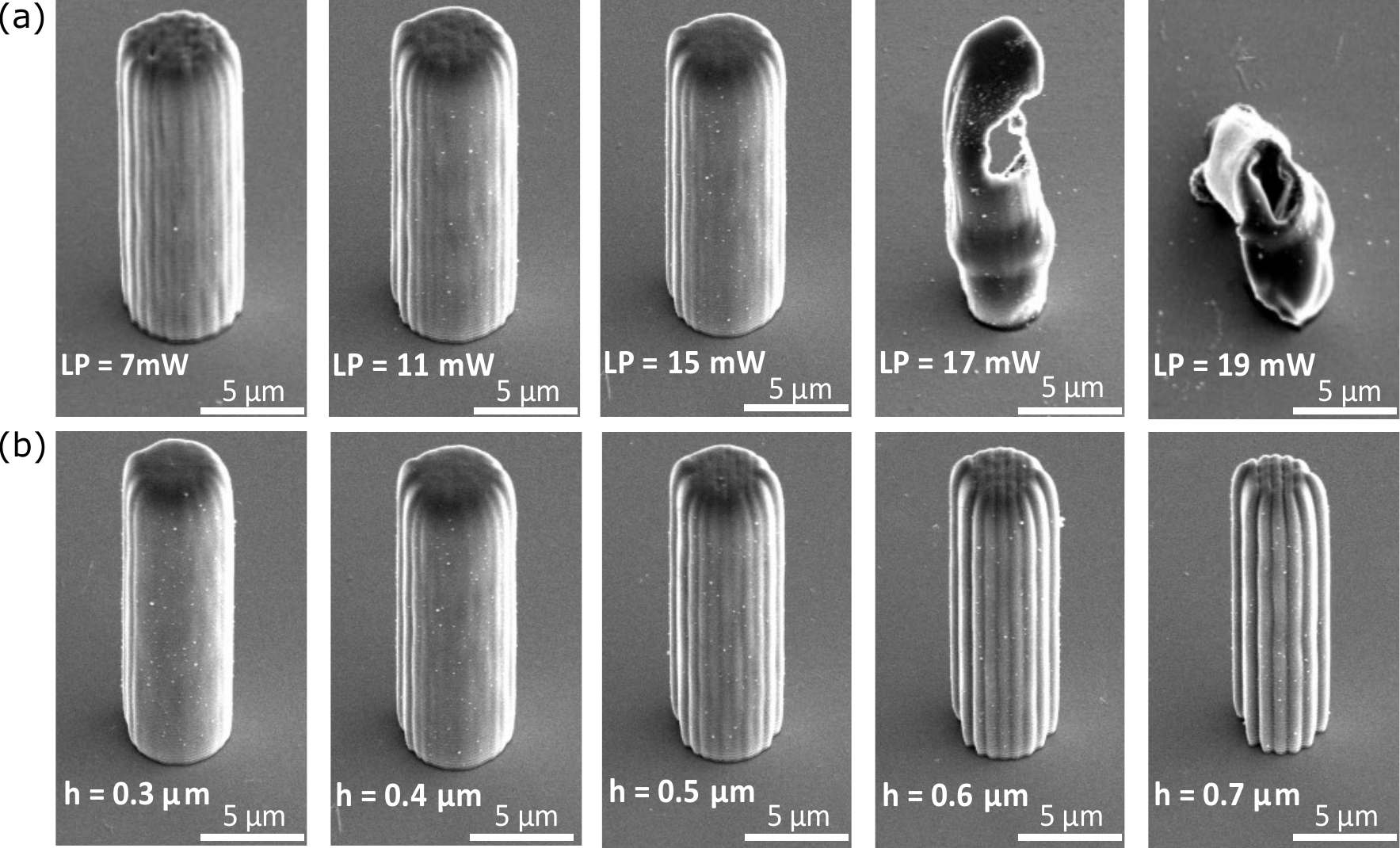}
	\caption{Dynamic power range characterization of waveguide cores printed via TPP using the IP-S photo-resist. Image taken with permission from~\cite{Grabulosa2022}.  
		(a) SEM micrograph of pilars, printed to reassemble the cores of waveguides, with 20~$\mu$m height and $d=5~\mu$m, with laser power LP $\in \{7, \dots, 19 \}$~mW, using hatching $h=0.4~\mu$m and slicing distance $s=1~\mu$m.
		(b) Impact of hatching distance $h\in \{0.3:0.1:0.7\}~\mu$m, with fixed LP = 15~mW and $s=1~\mu$m.
	}
	\label{fig:Fig_Dynamicpowerrange}
\end{figure}  

Figure \ref{fig:Fig_Dynamicpowerrange} (a-b) depicts the experimental optimization of the dynamic power range of the liquid negative-tone IP-S photo-resist, with $n\approx$ 1.51 when fully TPP-polymerized~\cite{Gissibl2017,Li2019} and using a 25X magnification NA $=0.8$ microscope objective for writing.
We printed, on a fused silica substrate, a set of five free-standing pillars to emulate waveguide cores with 20~$\mu$m height and diameter $d~=~5~\mu$m using a range of TPP laser power LP $\in \{7, \dots, 19 \}$~mW and hatching distances $h\in \{0.3:0.1:0.7\}~\mu$m.
As globally fixed parameters in all our fabrications we use a scanning speed of 10~mm/s and a slicing distance of $s=1~\mu$m.
The scanning electron microscopy (SEM) micrograph in Fig. \ref{fig:Fig_Dynamicpowerrange} (a) shows the effect of gradually modifying the LP with a hatching distance constant $h=0.4~\mu$m. 
Structures printed with LP~=~7~mW and LP~=~11~mW have ondulated surfaces, whereas when increasing the laser power to LP~=~15~mW results in larger TPP voxels and therefore smoother surfaces. 
Exceeding LP~=~15~mW leads to overpolymerization of the IP-S photo-resist (see two last micrographs of Fig \ref{fig:Fig_Dynamicpowerrange} (a)). 
We therefore select LP~=~15~mW and proceed to optimize the second fabrication parameter by scanning the hatching distance from $h\in \{0.3:0.1:0.7\}~\mu$m, and Fig. \ref{fig:Fig_Dynamicpowerrange} (b) shows the results. 
We found that for $h=0.3~\mu$m results are not always reproducible since smaller hatching distance increases local exposure dose $D$ and hence moves the process above the available power range.

\subsection{\label{subsec:OPP}One-photon polymerization}

\begin{figure}[h!]
	\centering
	\includegraphics[width=0.90\linewidth]{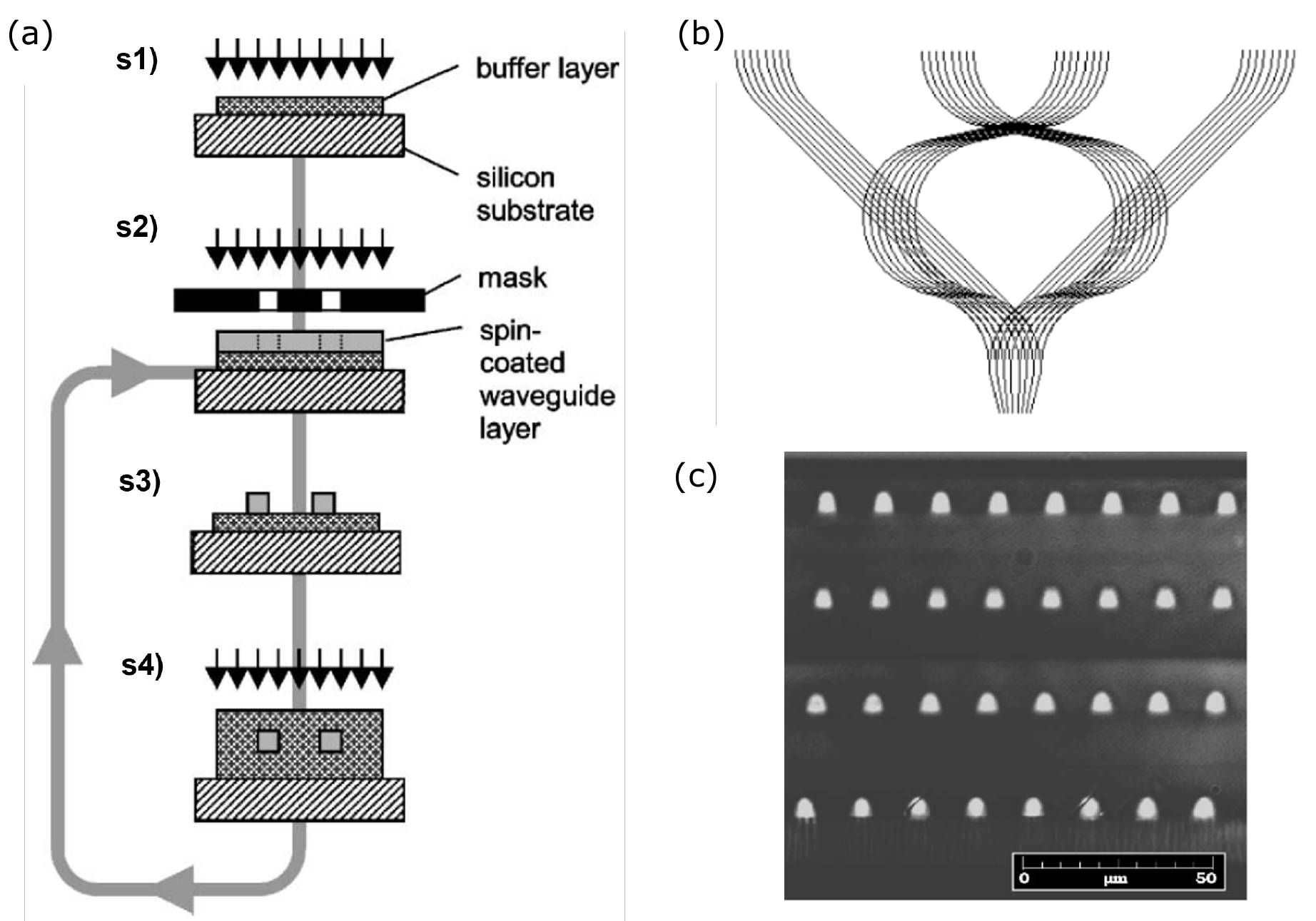}
	\caption{Multilayer 3D waveguide fabrication using OPP. Image taken with permission from~\cite{streppel2003new}. (a) Schematic diagram of the fabrication sequence for the stacking waveguide using spin coating and simple direct UV photolithography curing (s1); UV irradiation of the waveguides using a mask (s2); development (s3); UV irradiation of the cladding (s4). (b) Layout of the 3D interconnect polymer structure with an array of 4x8 waveguides. (c) Cross-section microscope optical image of 4x8 stack waveguides.
	}
	\label{fig:Uv_3D}
\end{figure}

One-photon polymerization is widely used to process thin material layers in the current 2D photo-lithography technology used for electronic semiconductor ICs.
The process is based on the exposure of a photosensitive resin, usually at the UV range, through a photo-mask including specific design patterns.
Repeating this process layer-by-layer is possible to process and stack different thin material layers and fabricate 3D structures \cite{garner1999vertically}.
For highly structured patterns like SD memory cards, this has led to ICs with up 100 or more circuit layers \cite{Boahen2022}.
However, such stacking of layers created via a generically 2D fabrication concept has several severe drawbacks.
For one, it requires to precisely align the photo-mask multiple times in each photo-lithographic step, which is challenging and time-consuming.
Secondly, one of the strongest features of 2D lithography is its economic appeal.
Between each layer, each of the process step have to be repeated in a loop-like manner.
A process where the entire IC's volume is created during few of such process steps will potentially have the upper hand economically speaking.
Still, such stacked 2D lithography has also been used of complex 3D photonic integration, c.f. Fig.~\ref{fig:Uv_3D}. 

Just as with TPP, the refractive index of the polymerized resin is a function of the optical exposure does $D$ \cite{Gissibl2017, Dottermusch2019, Schmid2019, TuingZukauskas2015}.
However, in OPP the refractive index of the resin is modified for substantially larger volumes, and in particular volumes outside the intended plane of exposure do strongly accumulate unintended irradiation doses.
It is therefore a formidable challenge to precisely control a 3D refractive index distribution, i.e. a volume hologram, with high spatial resolution.
OPP is therefore better suited for simultaneous polymerization of, either, large areas like in classical 2D lithography, or for large uniform volumes. 

\subsection{\label{subsec:TPPOPP}$Flash$-TPP: combining one- and two-photon polymerization for photonic integration}

\begin{figure}[h!]
	\centering
	\includegraphics[width=0.8\linewidth]{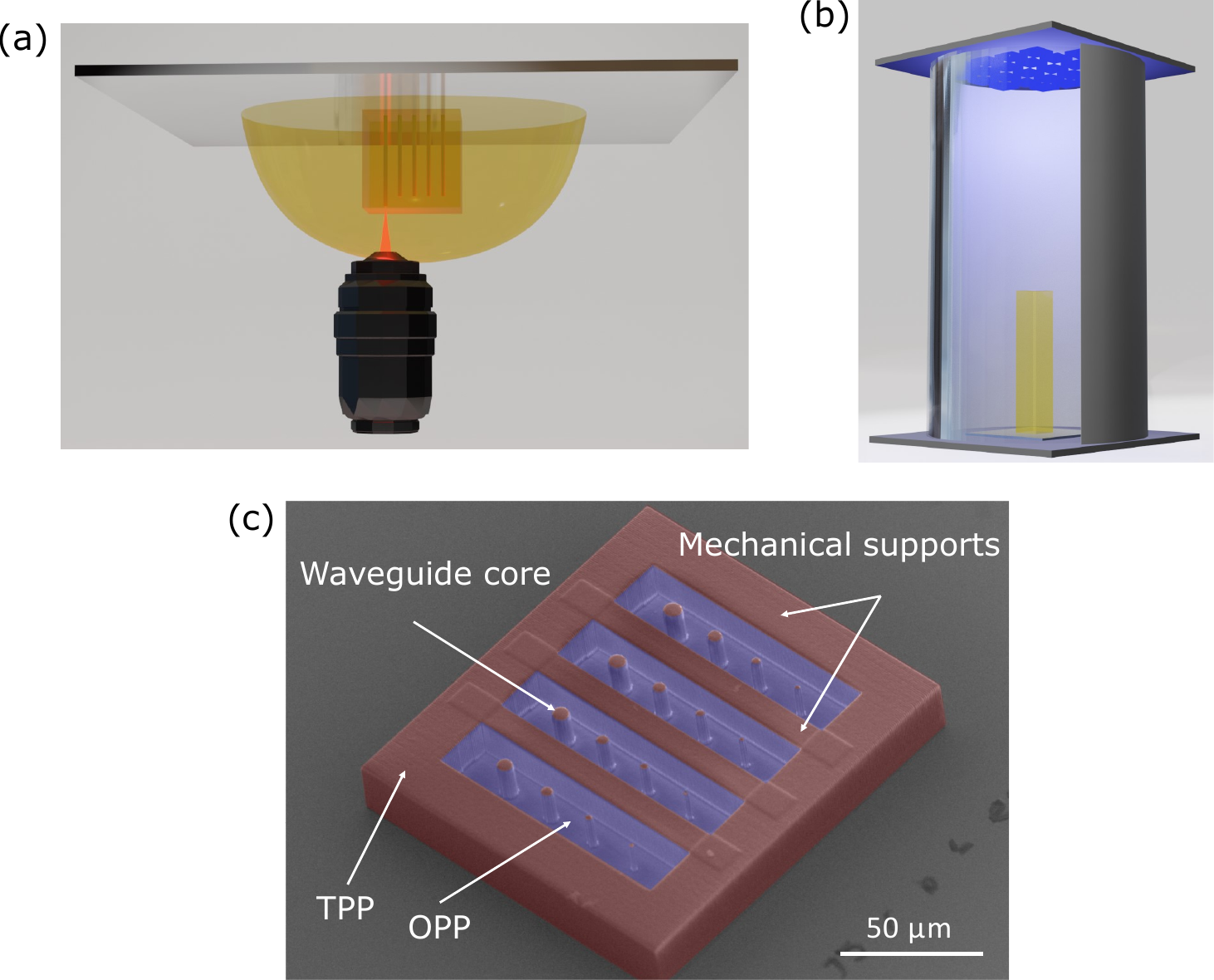}
	\caption{\emph{Flash}-TPP printing concept for 3D integrated photonics.
		Image taken with permission from~\cite{Grabulosa2022}.
		(a) Classical 'dip-in' process for the DLW-TPP fabrication of 3D photonic waveguides.
		(b) UV chamber that polymerizes the unexposed regions of the 3D structure via OPP.
		(c) SEM micrograph of a 3D-printed cuboid cross-section embedding 16 photonic waveguides.  The waveguide cores (mechanical supports) are printed with large (small) hatching distances, which defines the resolution of each component of the 3D photonic circuit.
		Red colour represents regions polymerized via TPP, while blue colour regions via OPP. 
	}
	\label{fig:Fig_TPPvsOPP}
\end{figure}  

One can combine one- and two-photon polymerization as an hybrid configuration to accelerate the fabrication of 3D photonic chips.
Several approaches combining UV lithography with DLW-TPP have been previously demonstrated in \cite{Eschenbaum2013} and \cite{Lim2018} for the fabrication of high resolution 3D optical microcomponents.
However, those methodologies require the polymerization of multiple photo-resists in two separated fabrication steps and become time-consuming if used for 3D fabrication due to the layer-by-layer approach.  

We demonstrated a novel lithographic strategy that combines OPP and TPP, \emph{flash}-TPP~\cite{Grabulosa2022}, where we combine high resolution and quality TPP with unstructured and uniform OPP in order to accelerate the fabrication process by one order of magnitude when compared to using TPP-only.
Importantly, the concept only requires a single resin and adding the OPP step does not add additional development and washing steps.
In \emph{flash}-TPP, TPP and OPP are used for the fabrication of the different sections of a photonic circuit, Fig.~\ref{fig:Fig_TPPvsOPP} illustrates the working principle, here for the liquid negative-tone IP-S photo-resist.
Waveguide cores accommodate the large majority of an optical signal's electromagnetic field, hence cores are printed via TPP with a precisely optimized laser power and fine resolution in the $(x,y)$-plane, i.e. small hatching distance.
This ensures smooth core-cladding interfaces and hence low propagation losses.
Mechanical supports, i.e. surfaces that define the outer limits of the volumetric circuit, are printed with larger hatching distance and high LP.

Figure \ref{fig:Fig_TPPvsOPP} (a) depicts the typically 'dip-in' DLW-TPP printing procedure.
After development, the photonic circuit is transferred to a UV chamber, c.f. Fig. \ref{fig:Fig_TPPvsOPP} (b), and the OPP dosage $D$ of the 3D circuit's volume is controlled via the duration of the UV exposure, through which we tailor the refractive index of the waveguides' cladding $n_{\textrm{cladding}}$ and hence $\Delta{n}$.
The SEM micrograph from Fig. \ref{fig:Fig_TPPvsOPP} (c) shows the cross-section of an exemplary 3D photonic chip fabricated via \emph{flash}-TPP consisting of a cuboid integrating 16 waveguides.
The cores and mechanical supports, printed via TPP, are highlighted in red region, while the cladding volume, polymerized via OPP, is highlighted in blue. 

\begin{figure}[h!]
	\centering
	\includegraphics[width=0.90\linewidth]{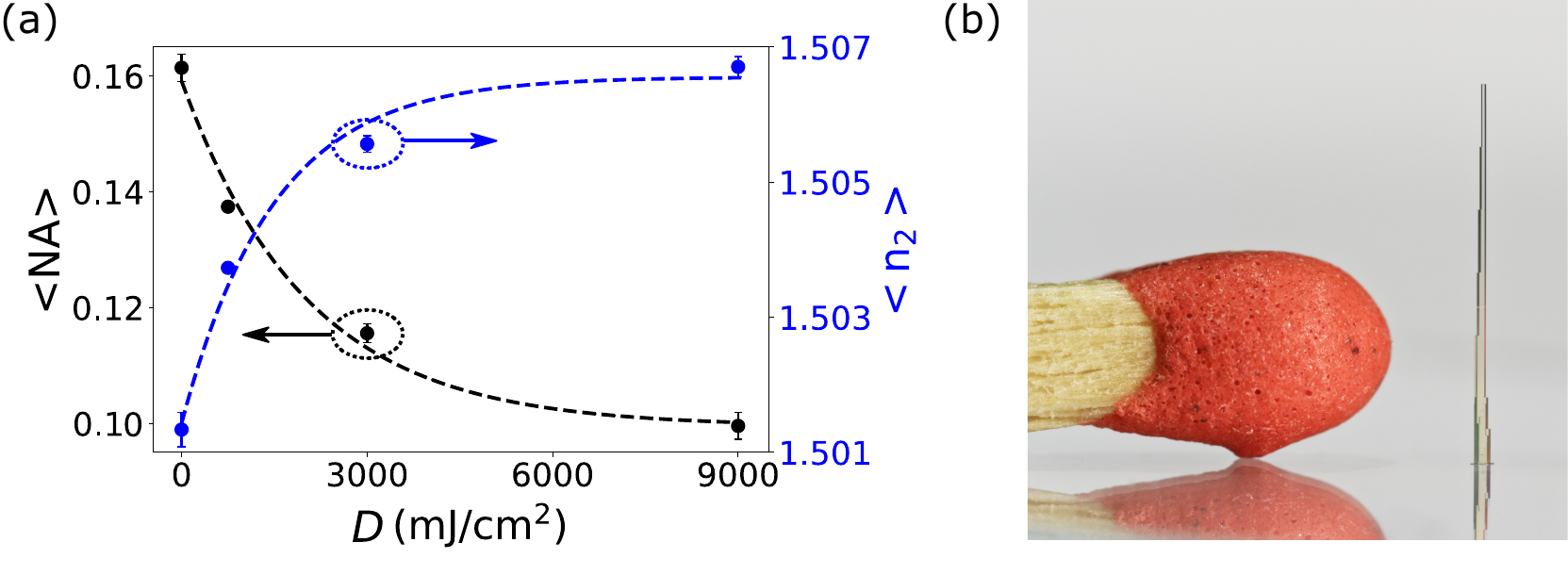}
	\caption{Optical performance of waveguides printed via \emph{flash}-TPP. Image taken with permission from~\cite{Grabulosa2022}.
		(a) Average numerical aperture $<$NA$>$ and cladding's refractive index $<n_2>$ over OPP dose $D$ of photonic waveguides printed via \emph{flash}-TPP. 
		The $<$NA$>$ ($<n_2>$) decreases (increases) over $D$, meaning that we can control the degree of polymerization of the cladding via the dosage of UV light.
		(b) Macroscopic structure scaled to a match that integrates waveguides with heights ranging from 0.1 to 6~mm.  
	}
	\label{fig:Fig_OPP_macro}
\end{figure}

Via \emph{flash}-TPP, we fabricated photonic waveguides with a refractive index contrast between core and cladding in the order of $\Delta$n $\approx$ 5$\cdot$10$^{-3}$~\cite{Grabulosa2022}.
Figure \ref{fig:Fig_OPP_macro} (a) shows the evolution of the the average numerical aperture $<$NA$>$ and refractive index of the cladding $<n_{\textrm{cladding}}>$ polymerized via OPP versus $D$.
We used UV exposure doses $D$ of 0, 750, 3000 and 9000 mJ/cm$^2$, respectively.
Assuming a constant $n_{\textrm{core}}\approx{1.51}$, we can precisely control, both, $<$NA$>$ and $<n_{\textrm{cladding}}>$.
Waveguides are single-mode for $d \leq 4.9~\mu$m, which are feasible to fabricate via standard DLW-TPP processes.
We obtained 1.3~dB/mm (0.26~dB) propagation (injection) losses for the fundamental LP$_{01}$ mode of waveguides printed via \emph{flash}-TPP.
Crucially, our 3D circuits did not degrade over time, and we evaluated the NA of waveguides under continuous operating condition across several months~\cite{Grabulosa2022}.
Overall, this demonstrates the reliability of the \emph{flash}-TPP lithography methodology for an ultra-fast, single-step and high performance fabrication of 3D photonic components. 

Printing via \emph{flash}-TPP consist in polymerizing only the sections vital for communication and mechanical integrity.
Importantly, the majority of a circuit's area or volume is not involved in either, and they can hence be rapidly fabricated via UV blanket exposure.
The printing times in \emph{flash}-TPP is therefore drastically reduced, and in particular cases also scales different with the circuit's size~\cite{Grabulosa2022}.
This agrees with our experience; \emph{flash}-TPP reduces the printing time to only 10$\%$ compared to only-TPP.
As an example, printing a large structure that integrates waveguides with heights ranging from 0.1 to 6~mm~\cite{Grabulosa2022}, shown in Fig. \ref{fig:Fig_OPP_macro} (b), requires $\sim$24 hours only using TPP but only $\sim$3 hours using \emph{flash}-TPP.

\section{\label{section:air_cladded}Air-cladded waveguides}

Polymer waveguides with an air cladding have a relatively large $\Delta{n}\approx0.5$ with $n_{\textrm{core}}=1.51$.
On the one hand, this leads to very strong confinement and a large $\textrm{NA}=1.13$, which enables very small bending radii of 25~$\mu$m (14~$\mu$m) at $\lambda=1550~$nm ($\lambda=650~$nm), and in turn dense photonic integration~\cite{Shane2011,Bahadori2019,Lapointe2020}.
The large $\Delta{n}$ makes fabricating single-mode waveguide circuits challenging.
To be single-mode, air-cladded waveguides have to have a core diameter $d\leq{1}~\mu$m ($d\leq{0.43}~\mu$m) at $\lambda=1550~$nm ($\lambda=650~$nm).
Printing waveguides with $d\leq{1}~\mu$m is possible \cite{Moughames2020}, and strongly confined photonic IC at $\lambda=1550~$nm are within reach.
For photonic 3D ICs close to the visible wavelength of light this remains a challenge.

\begin{figure}[h!]
	\centering
	\includegraphics[width=0.75\linewidth]{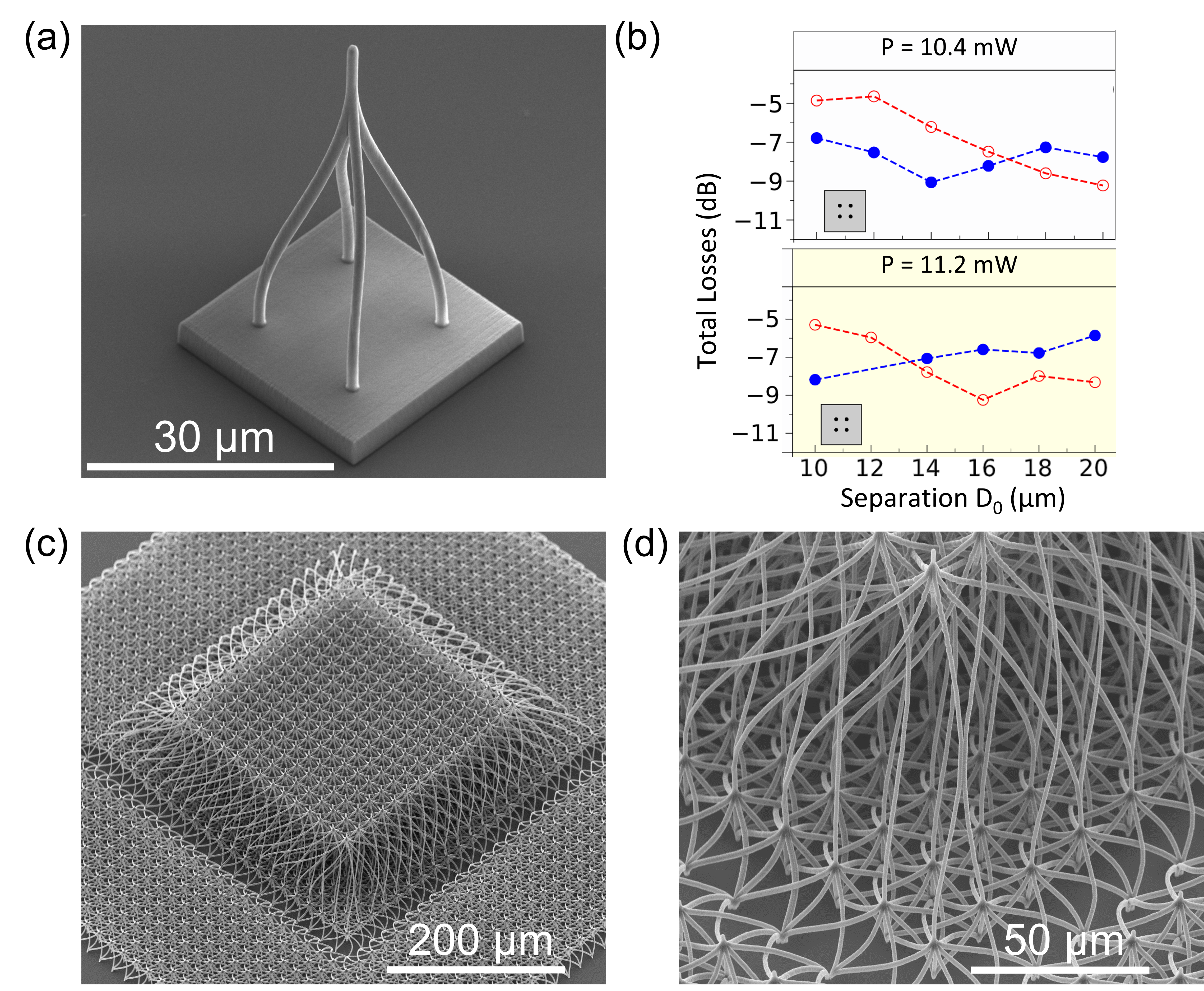}
	\caption {Air-cladded waveguides and couplers fabricated via DLW-TPP. 
		Image taken with permission from~\cite{Moughames2020,moughames20203d}.
		(a) 2x2 optical splitter/coupler with 1 input and 4 outputs with distance D$_{0}=16~\mu$m between waveguides, and 1.2~$\mu$m waveguide diameter~\cite{moughames20203d}.
		(b) Optical losses of 2x2 splitters/couplers as a function of the distance D$_{0}$ between waveguides, for hatching distances $h$~=~0.1~$\mu$m (in blue) and $h$~=~0.2~$\mu$m (in red). Data on top correspond to splitters/couplers written with laser power LP~=~10.4~mW, and data at the bottom correspond to splitters/couplers written with laser power LP~=~11.2~mW.
		(c) SEM micrographs of 3D-printed waveguides realizing parallel interconnects with high connectivity \cite{Moughames2020}.
		(d) Zoom-in of (c).
	}
	\label{fig:529o}
\end{figure}

Recently, 3D optical splitter/combiners based on air-cladded waveguides with a 1 to 4, 1 to 9 and 1 to 16 configuration were printed using TPP \cite{moughames20203d, moughames20213d}.
Figure \ref{fig:529o} (a) shows an SEM image of the 1 to 4 fractal splitter/coupler, with its optical characterization at $\lambda=632~$nm shown in Fig.~\ref{fig:529o} (b).
There, the distance between output ports was scanned  within the range $D_{0} \in [10, 12, ..., 20]~\mu$m while keeping their height constant at 52 $\mu$m.
Losses do not substantially increase for smaller distance between the output ports, which validates the estimated minimal bending radii given before.
Furthermore, this performance was evaluated for two different LP settings.
No clear difference can be seen between the two data-sets, and hence the printing power for air-cladded 3D polymer waveguides is not a critical parameter, as long one stays within the dynamic power range.

For large-scale network interconnect, Moughames \textit{et al.} demonstrated 3D parallel interconnects with high connectivity, shown in Figure \ref{fig:529o} (c), by cascading two layers of 1 to 9 splitters and spatially multiplexing an arrays of such 1 to 81 splitters to allows for an array of 15x15 input waveguides.
The entire circuits only occupies a volume of 460x460x300 $\mu m^3$, in which an interconnect for  225 inputs and 529 outputs is realized \cite{Moughames2020}.
Figure \ref{fig:529o} (d) shows a higher magnification of this interconnect.
Individual wavegudies have a low surface roughness, and the incorporated chirality of the fractal splitters/couplers avoids intersections of individual waveguides.

\section{\label{section:GRIN_waveguides}Step and graded index waveguides}

Based on the previous discussed concepts and fabrication technologies, we addressed step- (STIN) and graded-index (GRIN) waveguides. 
In STIN waveguides, the refractive index of the waveguide's core is constant, while for GRIN waveguides it is a function of the radial distance to the core's center.
Usually, GRIN waveguides follow a parabolic refractive index distribution.
For the STIN waveguides, all bound rays propagate at angles within the total internal reflection condition $\theta_{\textrm{c}}$ at any position in the core cross-section, while for GRIN waveguides, the range of angles varies with position~\cite{Snyder1983}.

We proposed a single-step additive fabrication technique, (3+1)D printing~\cite{Porte2021}, by which we spatially modify the refractive index of a single resin over the TPP exposure dose during fabrication.
Using the (3+1)D-printing concept, we constructed volume holograms and photonic waveguides with, both, STIN and GRIN profiles in a single-step, single-material fabrication with a commercially available process.
This demonstrates the versatility of the 3D photonic integration approach based on DLW; optical manipulation based on integrated and monolithic 3D structures can either rely on discrete components, i.e. waveguides, or leverage continuous manipulations of free optical propagation, i.e. holograms~\cite{Porte2021}.
Both schemes can be exploited on the same photonic IC and be realized using the same fabrication concept and during the same fabrication step.
We used the negative tone IP-Dip resin ($n\approx$ 1.547)~\cite{Schmid2019} and a 63X magnification NA = 1.4 microscope objective, c.f. Fig.~\ref{fig:Fig_TPPvsOPP} (a).

\begin{figure}[h!]
	\centering
	\includegraphics[width=0.65\linewidth]{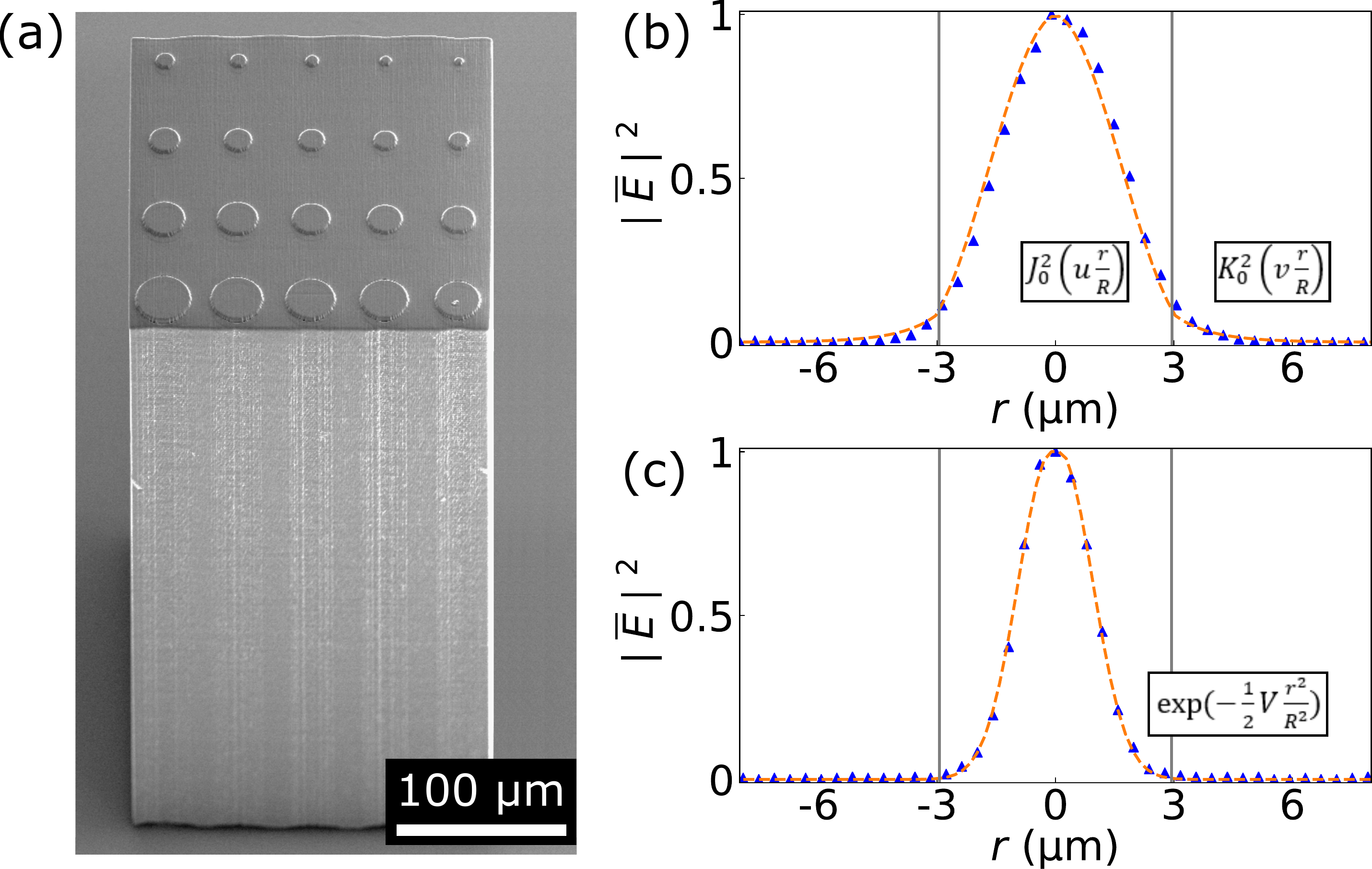}
	\caption{Step- (STIN) and graded-index (GRIN) waveguides fabricated via (3+1)D-printing. Image taken with permission from~\cite{Porte2021}.
		(a) SEM micrograph of an exemplary 3D-printed cuboid integrating 20 STIN waveguides of 300~$\mu$m heigh. Waveguide cores (cladding) are printed via TPP with high (low) laser power, which ensures a refractive index contrast $\Delta$n $\approx$ 2.4$\cdot$10$^{-3}$. 
		Panels (b) and (c) depict the output intensities (triangles) and fundamental LP$_{01}$ mode fits (dashed lines) of a 3~$\mu$m radius STIN and GRIN waveguide, respectively.
	}
	\label{fig:Fig_3+1D}
\end{figure}

The SEM micrograph of Fig.~\ref{fig:Fig_3+1D} (a) shows an exemplary cuboid embedding 20 STIN waveguides fabricated via (3+1)D-printing.
Contrary to \emph{flash}-TPP, in (3+1)D-printing all the 3D photonic chip volume is fabricated via TPP-only.
The refractive index contrast $\Delta{n}$ between core-cladding waveguides is achieved from the control over the TPP dosage $D$ for individual writing voxels.
For a higher (lower) refractive index as needed for the waveguide cores (claddings), one requires an accordingly higher (lower) LP, i.e. $D$.
STIN waveguides result from a constant LP all across their core, while for GRIN waveguides the writing power changes from high to low following a parabolic profile.

To evaluate the optical performance, we fitted the experimental output intensities for diameters $d$ below the cut-off condition of the second propagation mode.
The output intensity of the LP$_{01}$ mode of a STIN waveguides is described by $J_{0}^2(u \frac{r}{R})$ for $|$ r $|$ $<$ $R$ and $K_{0}^2(v \frac{r}{R})$ for $|$ r $|$ $>$ $R$, while for GRIN waveguides is given by an infinite parabolic refractive index profile as $\exp{-\frac{1}{2} V \frac{r^2}{R^2}}$~\cite{Snyder1983}.
Figure~\ref{fig:Fig_3+1D} (b-c) depicts the fit of fundamental LP$_{01}$ mode to the normalized output of STIN and GRIN waveguides with radius $R~=~3~\mu$m, respectively.
Considering the refractive index of the core constant ($n_\textrm{core}\approx$ 1.547), we obtained an averaged numerical aperture $<$NA$>$ = 0.08~$\pm$~0.01 (i.e. $n_\textrm{core}$~=~$n_\textrm{cladding}$~+~2.4~$\cdot$~10$^{-3}$) for STIN and of $<$NA$>$ = 0.18~$\pm$~0.02 for GRIN waveguides.
As expected, the core-confinement of GRIN waveguides is significantly higher than for STIN waveguides due to the inner core refractive index modification, which offers a crucial advantage for photonic integration schemes~\cite{Moughames2020}.

As seen, STIN waveguides with a polymer cladding have a refractive index contrast in the order of  $\Delta{n}\approx$ 2.4$\cdot$10$^{-3}$, with low NA $\approx$ 0.12.
Contrary than for air-cladded waveguides, this leads to large bending radii of 15~mm (7~mm) at $\lambda=1550~$nm ($\lambda=650~$nm), and in turn dense photonic integration is much more challenging for STIN waveguides.
However, the low $\Delta{n}$ allows to have single-mode propagation for waveguide diameters $d\leq{9.8}~\mu$m ($d\leq{4.2}~\mu$m) at $\lambda=1550~$nm ($\lambda=650~$nm), which is standard with the current DLW-TPP fabrication technology. 
Future efforts include combining polymer and air-cladded waveguides, taking the strengths of each configuration in a single platform, i.e. air cladding waveguides providing highly-densed photonic integration with their small bending radii, while STIN waveguides serving as tools for single-mode propagation with large waveguides diameters over wide distances.   

\section{\emph{Flash}-TPP printed waveguides}

Recently, we demonstrated the fabrication of large scale 3D integrated photonic components via \emph{flash}-TPP.
Several features of \emph{flash}-TPP make it an enabling technology for integration of larger circuits.
Of primary importance is the substantial accelerated fabrication; without, fabrication of larger integrated circuits would quickly approach timescales beyond 24h~\cite{Grabulosa2022}.
Based on this approach, we demonstrated long (6~mm) single-mode waveguides, and we achieved exceptionally low injection ($\approx0.26~$dB) and propagation ($\approx1.3~$dB/mm) losses~\cite{Grabulosa2022}. 

Next as the demonstration of optical splitters and combiners based on this concept.
These are the backbone of any photonic IC, and 3D integration enables interesting alternatives for creating 1 to M optical couplers without using sensitive optical interference units~\cite{Soldano1995}.
In 3D, 1 to M optical couplers can simply be realized by arranging numerous output waveguides around the input waveguide, something impossible to realize in a purely 2D integration setting.
We demonstrated broadband 1 to M splitters leveraging adiabatic coupling~\cite{Grabulosa2023,Grabulosa2023b}.
Adiabatic coupling achieves low-loss single-mode optical transfer from 1 to M waveguides through evanescent waves, where the optical mode adiabatically leaks from a tapered core of an input waveguide towards the cladding into  inversely-tapered cores of the output waveguides~\cite{Spillane2002,Collot1993}.
All the previous studies consider the 2D case of only one to one adiabatic coupling between optical components~\cite{Tiecke2015}. 

\begin{figure}[h!]
	\centering
	\includegraphics[width=0.6\linewidth]{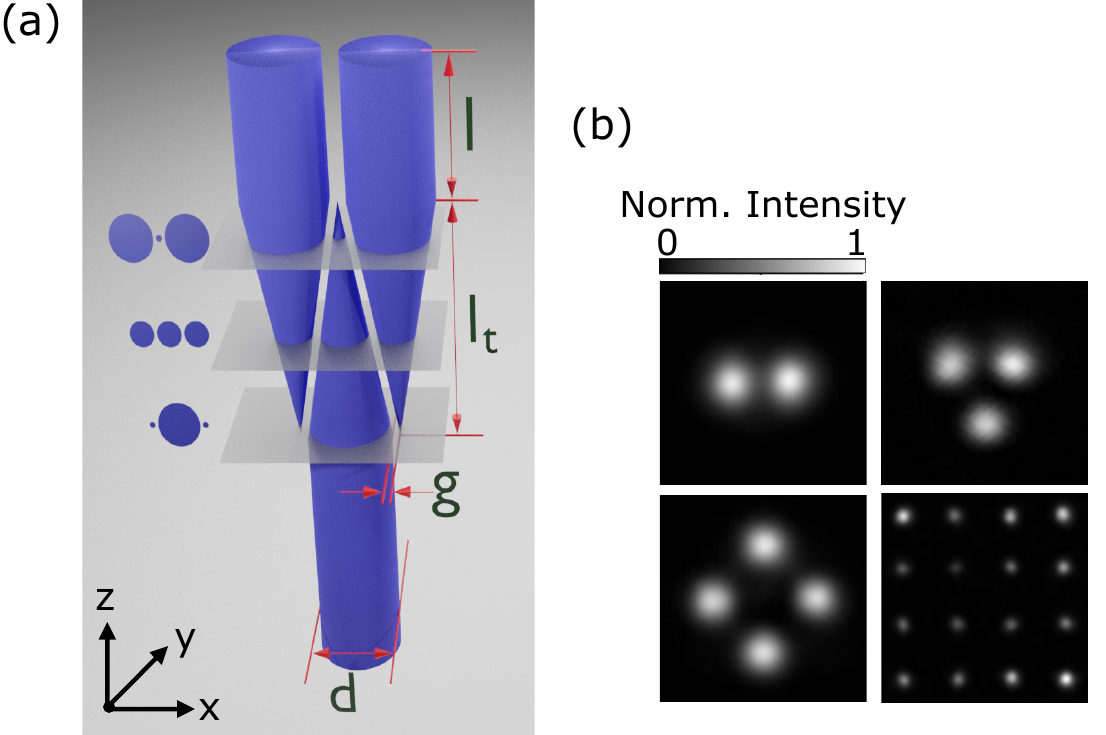}
	\caption{Adiabatic 1 to M broadband-scalable couplers fabricated via \emph{flash}-TPP.
		Image taken with permission from~\cite{Grabulosa2023}.
		(a) Design of the 1 to 2 adiabatic couplers printed via \emph{flash}-TPP. The same tapering strategy can be applied to higher-order couplers, i.e. 1 to 3 and 1 to 4 couplers. 
		(b) Output intensity profiles of the 1 to 2, 3 and 4 adiabatic couplers. The last output intensity corresponds to a cascaded 1 to 16 adiabatic coupler. 
	}
	\label{fig:Fig_adiabatictapers}
\end{figure}

In our work, we showed efficient single-mode adiabatic transfer with 1 input and up to 4 outputs via a single component.
Figure~\ref{fig:Fig_adiabatictapers} (a) illustrates the design for the exemplary case of a 1 to 2 adiabatic couplers.
The waveguide's circular core cross-section continuously changes as a function of propagation direction $z$.
The originally circular core is reduced in size exclusively along the directions where an output waveguide is located; the core is essentially cut along plane surfaces.
These cut-planes move towards the input core's center during the taper-length $l_{\textrm{t}}$ at equal rate $d/l_{\textrm{t}}$ along the $(x,y)$-plane in order to match their relative effective modal indices~\cite{Snyder1983}.
Output waveguides follow exactly the same concept, yet in an inverted direction.
We separated in and output waveguides via gap $g$ and studied the evanescence coupling efficiency between coupled waveguides~\cite{Grabulosa2023}.
The same tapering strategy was applied to 1 to 3 and 1 to 4 as depicted in the output intensity profiles from Fig.~\ref{fig:Fig_adiabatictapers} (b).

We obtained record optical coupling losses of 0.06~dB for the optimal case of 1 to 2 adiabatic couplers, with a difference between the two outputs intensities of only $\sim3.4~\%$.
We furthermore demonstrated broadband functionality from 520~nm to 980~nm during which losses remain below 2~dB~\cite{Grabulosa2023}.
Importantly, these adiabatic couplers can be cascaded in order to exponentially increase the number of M outputs, c.f. Fig.~\ref{fig:529o} (c).
We arranged a double-layer of 1 to 4 adiabatic couplers and the resulting 1 to 16 single-mode output intensities can be seen in the last diagram of Fig.~\ref{fig:Fig_adiabatictapers} (b).
Importantly, the global losses of the entire device is only 1~dB~, and the entire circuit was realized within $(0.08\times0.08\times1.5)~$mm$^3$~\cite{Grabulosa2023}. 

\section{\label{section:photonics_networks}Towards a scalable and CMOS compatible integration of photonic networks}

High-density photonic integration requires the interconnection of several photonic platforms.
Most of the current photonic devices are based on silicon-on-insulator (SOI) and CMOS technology.
Combining the strength of multiple photonic and electronic systems in one hybrid and multi-chip platform can result in the diversification of specific computing tasks while increasing the overall performance.

\begin{figure}[h!]
	\centering
	\includegraphics[width=1\linewidth]{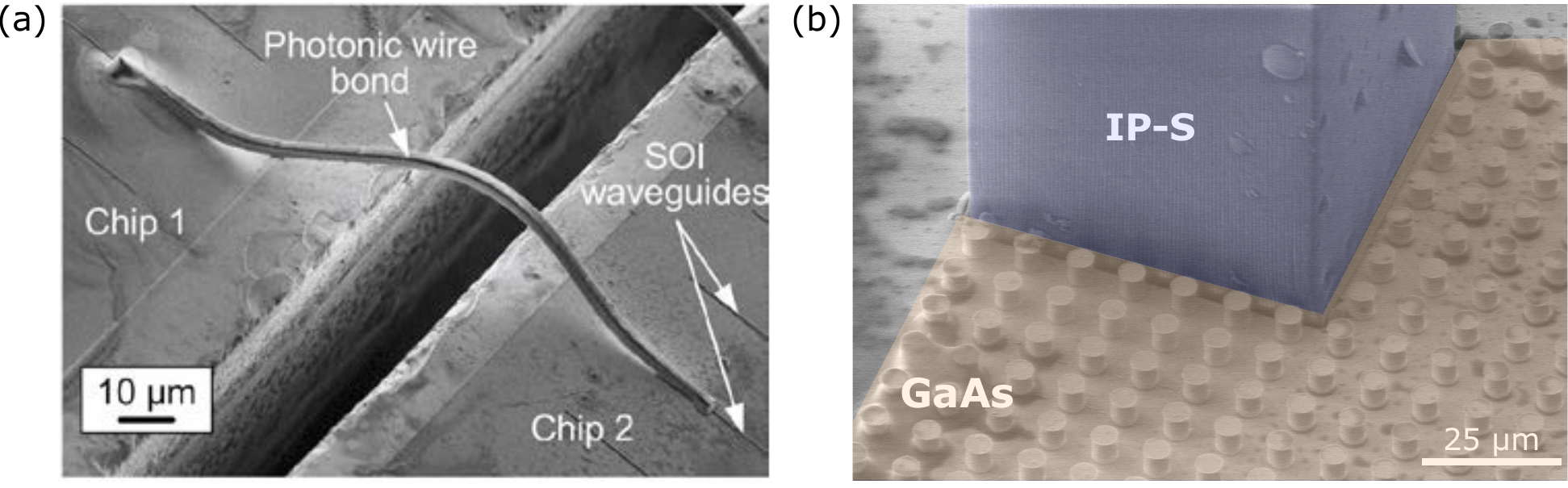}
	\caption{Polymer-based 3D printing and CMOS technology compatibility. 
		(a) Chip-to-chip photonic wire bonding concept. A 3D polymer waveguide fabricated via DLW-TPP connects two SOI waveguides sitting on distant CMOS chips. SEM image taken with permission from ~\cite{Lindenmann2012}. 
		(b) SEM micrograph of and exemplary 3D-cuboid integrating a cascaded 1 to 16 adiabatic couplers printed via \emph{flash}-TPP on top of a quantum dot micropillar laser array. 
	}
	\label{fig:Fig_CMOS}
\end{figure}

A versatile fabrication technology with low-losses is of vital importance for the scalability of free-form as well as integrated optical interconnects in three-dimensions.
The polymer-based 3D printing technology based on DLW-TPP is excellently suited to address these challanges, and several proof-of-concept studies have been realized~\cite{Tiecke2015,Nesic2019,Saeed2020}.
Figure~\ref{fig:Fig_CMOS} (a) shows photonic wire-bonding, realising a 3D photonic waveguide forming a point to point communication for a chip-to-chip connection between SOI chips hosting individual waveguides.
The photonic wire-bond was fabricated via DLW-TPP using the negative-tone MicroChem SU-8 2075 photo-resist ($n\approx$ 1.51 at 1550~nm)~\cite{Lindenmann2012}, and it connected two SOI waveguides separated a distance of 100~$\mu$m on different CMOS chips.
This demonstrated for the fist time the basic viability of TPP-based 3D printing as a tool for CMOS compatible, wafer-scale as well as chip-to-chip connections. 

A major challenge of the polymer-based 3D fabrication and the CMOS technology is the interaction of the CMOS substrate with the photo-resist during the TPP printing process.
In a standard fabrication setting, the interaction between the fs-pulsed laser and the glass substrate is negligible since the substrate material, i.e. fused silica, is transparent at the wavelength of the fs-laser (780~nm), and low specular reflection.
However, the CMOS technology is based on 2D stacking of multiple thin layers of semiconductor materials such as GaAs, InP or Silicon.
These often have a bandgap energy below that of the writing laser, and in that case printing through the semiconductor substrate is impossible; only the 'dip-in' concept is therefore a viable general approach for fabricating 3D photonic integrated circuits directly on top of a CMOS substrate based on DLW-TPP. 
Another challenge is the higher specular reflection, as these semiconductor materials have a higher refractive index.
The resulting optical reflection of the fs-laser laser at the semiconductor substrate leads to a overpolymerization of the photo-resist if not compensated for.
The LP therefore needs to be continuously adjusted at the vicinity of the CMOS/photonic circuit interface in order to achieve the intended degree of polymerization of the photo-resist.
A further requirement is the precise alignment of the 3D photonic chip with the semiconductor device patterned on the CMOS substrate.

Figure~\ref{fig:Fig_CMOS} (b) depicts an exemplary 3D-printed cuboid integrating a cascaded 1 to 16 adiabatic coupler (cf. Fig.~\ref{fig:Fig_adiabatictapers} (b)) printed via \emph{flash}-TPP on top of a semiconductor substrate integrating quantum dot micropillar laser arrays.
Each of the micropillar lasers consists of a cylindrical microcavity (a vertical arrangement of highly reflective distributed Bragg reflectors (DBR) alternating Al(Ga)As and GaAs mirror pairs) sandwiching a central gain section based on InGaAs self-assembled quantum dots (QDs).
Further details about the fabrication and optical properties of the quantum dot micropillars laser arrays from Fig. \ref{fig:Fig_CMOS} (b) can be found in \cite{Reitzenstein2007,Heuser2018,Reitzenstein_2010}.
We used IP-S photo-resist for the fabrication, with a lower laser power LP~=~6.5~mW (compared to the previously LP~=~15~mW) in order to avoid microexplosions of the photo-resist at the semiconductor-polymer interface during TPP printing.
After development, the 3D photonic chip is then polymerized via OPP with a exposure dose $D~=$~3000~mJ/cm$^2$.
The SEM micrograph shows the perfectly aligned 3D photonic structure with the angle of the periodic GaAs micropillar array.
We checked the adherence of the polymer over time, and after a continuously observation over more than 4 months no deterioration has been found.
This confirms the reliability of integrating our 3D printing technology with CMOS-based micro-laser arrays.

\section{Conclusion}

Here, we present a review over our recent work addressing additive manufacturing towards future 3D photonic integration of optical components that is CMOS compatible. 
Based on one- and two-photon polymerization processes combined with direct-laser writing systems, we demonstrated the fabrication of high performance individual photonic waveguides as well as scalabale optical splitters. 
All such 3D structures have been fabricated in our local FEMTO-ST RENATECH infrastructure. 

We demonstrated that using the commercial DLW-TPP Nanoscribe GmbH (Photonics Professional GT) system and the 'dip-in' DLW strategy, we are able to the construct, both, air- and polymer-claddded photonic waveguides. 
For air-cladded waveguides, we used a TPP-only, a single-step and single resin (IP-Dip resist).
A 3D IC comprising a network of fractal optical splitter with 225 input and 529 output waveguides only occupies a volume of 460x460x300 ${\mu}m^3$.  
Such air-cladded waveguide ICs are prime candidates for highly-dense photonic packaging thanks to their low bending-radii on 10s of $\mu{m}$ scale. 
For polymer-cladded waveguides, we presented two different strategies in which we 3D-printed the waveguide cores via TPP while achieving a precise control over the refractive index contrast $\Delta{n}$ via, (i), the adjustment of the fs-laser dose $D$ on an single-voxel level, i.e. (3+1)D-printing, and (ii), the duration of UV blanket exposure that determines the OPP dosage $D$ to fix the index of the cladding material for the entire photonic IC in a single shot, i.e. \emph{flash}-TPP.
Noteworthy, both fabrication concepts require a single procedure writing step and a single resin (IP-S resist).
Importantly, with \emph{flash}-TPP fabrication times are reduced by up to $\approx 90~\%$ compared to (3+1)D-printing thanks to the additional OPP process.  
Via \emph{flash}-TPP, we achieved polymer-cladded waveguides with refractive index contrast $\Delta{n}\approx$ 5$\cdot$10$^{-3}$, with low 1.3~dB/mm (0.26~dB) propagation (injection) losses while printing waveguides up to 6~mm heigh.
This allows to have single-mode propagation over large distances.
We demonstrated the fabrication, via \emph{flash}-TPP, of scalable-boadband couplers leveraging adiabatic transfer from 1 input up to 4 outputs. 
Using a tapered/inversely-tapered waveguide sequence, we achieved record 0.06~dB optical coupling losses with very symmetric splitting ratios.
We arranged a double-layer of 1 to 4 adiabatic couplers, resulting in a device with 16 single-mode outputs with only 1~dB global losses.

Importantly, we demonstrated the compatibility of our fabrication methodology based on DLW-TPP with CMOS substrates. 
As a proof-of-concept, we successfully 3D-printed our cascaded 1 to 16 adiabatic couplers on top of a CMOS substrate integrating GaAs quantum dot micropillar laser arrays. 
Preliminary characterization of these structures shows encouraging performance in terms of losses and stability.

Overall, in this review we have covered our novel 3D-printing technology, which represents a breakthrough with the potential to become a high-impact tool for the hybrid, highly-dense and hence compact packaging of, both, electronic and photonic devices.
The concepts opens several potential avenues for future exploration.
The combination of air- and polymer-cladded waveguides could enable dense integration with simultaneous precise control over optical signal properties such as mode number, polarization and phase.
As the concept leverages photo-polymerization, in principle the large-scale and exceptionally performing production facilities of CMOS electronic integration could be amended with 3D photonic integration capability.
Due to the excellent compatibility of standard photo-resins, the approach is largely agnostic to the underlying substrate.
In this it is more flexible than integrated silicon photonics, and fabricating additively on a already processed CMOS substrate removes many of the challenges compared to fabricating photonic ICs based on different process - such as DLW directly into bulk dielectrics followed by bonding to CMOS.

\section{Acknowledgment}

The authors would like to thank Stephan Reitzenstein for his contribution through fabricating the semiconductor laser sample used for producing the circuit shown in Fig. \ref{fig:Fig_CMOS} (b) and Erik Jung for the valuable help on the design of 3D waveguides. 
This work was partly supported by the french RENATECH network and its FEMTO-ST technological facility.
The authors acknowledge the support of the Region Bourgogne Franche-Comté.
This work was supported by the EUR EIPHI program (Contract No. ANR-17-EURE- 0002), by the Volkswagen Foundation (NeuroQNet II), by the French Investissements d’Avenir program, project ISITE-BFC (contract ANR-15-IDEX-03), by the European Union’s Horizon 2020 research and innovation programme under the Marie Skłodowska-Curie grant agreements No. 713694 (MULTIPLY).
	
\bibliography{bibliography}
\bibliographystyle{ieeetr}
	
\end{document}